# Neural Crystals

Sofia Karamintziou
*Information Technologies Institute*
*Centre for Research & Technology-Hellas*
Thermi, Thessaloniki, Greece
skaramintziou@iti.gr

Thanassis Mavropoulos
*Information Technologies Institute*
*Centre for Research & Technology-Hellas*
Thermi, Thessaloniki, Greece
mavrathan@iti.gr

Dimos Ntioudis
*Information Technologies Institute*
*Centre for Research & Technology-Hellas*
Thermi, Thessaloniki, Greece
ntdimos@iti.gr

Georgios Meditskos
*School of Informatics*
*Aristotle University of Thessaloniki*
Thessaloniki, Greece
gmeditsk@csd.auth.gr

Stefanos Vrochidis
*Information Technologies Institute*
*Centre for Research & Technology-Hellas*
Thermi, Thessaloniki, Greece
stefanos@iti.gr

Ioannis (Yiannis) Kompatsiaris
*Information Technologies Institute*
*Centre for Research & Technology-Hellas*
Thermi, Thessaloniki, Greece
ikom@iti.gr

*Abstract*—We face up to the challenge of explainability in Multimodal Artificial Intelligence (MMAI). At the nexus of neuroscience-inspired and quantum computing, interpretable and transparent spin-geometrical neural architectures for early fusion of large-scale, heterogeneous, graph-structured data are envisioned, harnessing recent evidence for relativistic quantum neural coding of (co-)behavioral states in the self-organizing brain, under competitive, multidimensional dynamics. The designs draw on a self-dual classical description – via special Clifford-Lipschitz operations – of spinorial quantum states within registers of at most 16 qubits for efficient encoding of exponentially large neural structures. Formally 'trained', Lorentz neural architectures with precisely one lateral layer of exclusively inhibitory interneurons accounting for *anti-modalities*, as well as their co-architectures with intra-layer connections are highlighted. The approach accommodates the fusion of up to 16 time-invariant interconnected (anti-)modalities and the crystallization of latent multidimensional patterns. Comprehensive insights are expected to be gained through applications to Multimodal Big Data, under diverse real-world scenarios.

## I. Introduction

Advantages of quantum neural networks over their classical counterparts are being intensively investigated, as part of the nascent field at the intersection of machine learning and quantum information science [1]: quantum AI. It is hoped that, in addition to significant advancements in speed and capacity, quantum computing will offer a richer framework than classical computing for deep learning of complex representations of data [2], all the more so for geometric deep learning [3, 4]. At the same time, self-organizing artificial neural networks accounting for *antagonistic* interactions between excitatory neurons and inhibitory interneurons [5, 6] are emerging as a promising direction in *artificial general intelligence* as opposed to less biologically plausible deep learning networks. There are good reasons to believe that these developments may significantly benefit from concurrent critical advances in our understanding of multidimensional, brainwide activity by virtue of emergent large-scale neural recording technologies [7] and theoretical insights derived thereof [8, 9].

A valuable byproduct of this line of research may ultimately be to address major open challenges in applying AI across the complex network sciences, in the Big-Data era: the improvement of explainability and generalization ability of classical deep neural networks [10, 11], ideally moving from 'black-box' to *'glass-box'*, human-in-the-loop models [12]; and the search for novel neural architectures optimally requiring minimal or no training [13, 14], including quantum neural architectures [15].

In light of recent evidence for relativistic quantum neural coding of (co-)behavioral states in the self-organizing brain and an early crystallization of pertinent multidimensional synaptic (co-)architectures [9], we discuss formally 'trained', 'glass-box' neural architectures or, in short, *neural crystals*. Neural crystals are envisioned to allow (i) early fusion of multiple disparate, but interconnected datasets [16], namely large-scale, heterogeneous, graph-structured data, under scenarios where antagonistic interactions are inherently present; (ii) the recognition of underlying multidimensional patterns of representations. By geometrical reduction, the structures are steered towards efficient encoding of the information into a register of at most 16 qubits, the respective quantum states being classically describable via special Clifford-Lipschitz operations and self-duality under mirror supersymmetry. Emphasis is being placed on Lorentz neural crystals featuring a special multipartite graph topology of a low number of (non-hidden) layers.

## II. Asymmetric Symmetries

The aim is to construct quantum-classical neural structures with topology given by a weighted spin geometry, namely by a non-simple finite digraph, $\mathcal{D}_\gamma := \left( \mathcal{D}_\gamma \, ; \, \sigma \circ \kappa, \, \tau \circ \kappa \right)$, where $\mathcal{D}_\gamma := \left( \mathcal{D}, \gamma \right)$ denotes the weighted digraph with trivial structure associated to $\mathcal{D}_\gamma$. In particular, in the digraph $\mathcal{D} := \left( \mathcal{V}, \mathcal{E} \right)$ we let $\mathcal{V}$ be the set of its vertices and $\mathcal{E} \subseteq \mathcal{V} \times \mathcal{V}$ be a reflexive relation on $\mathcal{V}$. For the rest of technical and notational details

Funding by the European Union's Horizon Research and Innovation Actions, under grant agreements No. 101097036 (ONCOSCREEN), No. 101104777 (ONCODIR) and No. 101017558 (ALAMEDA), is gratefully acknowledged.

regarding the geometry $\mathcal{D}_\gamma$ and the self-organizing dynamics of the respective spinorial flow network, $\mathcal{F}$ (or the co-flow network, $\mathcal{F}^\perp$, obtained by functional duality), we refer the reader to [9]. Here, we consider it necessary to restate:

*Proposition:* Let $\vartheta: X \to \mathbb{R}$ denote a nondegenerate quadratic form on an $\mathbb{R}$-linear space $X$, with $dim_\mathbb{R} X = p + q < \infty$; therewith, $(X, \vartheta)$ is isomorphic to the pseudo-Euclidean space $\mathbb{R}^{p,q}$ with signature $(p,q)$ and positive index $p$. In $dim_\mathbb{R} X = p + q \leq 5$

$$\mathcal{O}(\delta) = \mathcal{O}^0_{p,q}.$$

That is, the null super-cone, $\mathcal{O}(\delta)$, of the quadratic norm, $\delta$, on the even Clifford algebra, $\mathfrak{C}^0_{p,q}$, of $(X, \vartheta)$ is *self-dual* ($\mathcal{W}$-*dual*) to the common boundary, $\mathcal{O}^0_{p,q}$, of the disjoint open sets $\Gamma^{0+}_{p,q}$ and $\Gamma^0_{p,q} \setminus \Gamma^{0+}_{p,q}$, $\Gamma^{0+}_{p,q}$ denoting the identity component of the special Clifford-Lipschitz group $\Gamma^0_{p,q}$. The statement is not true in $dim_\mathbb{R} X = p + q = 6$, with $p \neq 0$, where mirror super-symmetry breaks down.

*Translation:* The above proposition hints at a principle for (in)stability of a broad class of time-invariant, multidimensional self-organizing networked systems. Such (non-)equilibrium state ($\mathcal{W}$-*duality* or *mirror supersymmetry*) is economically effectuated by a bound on dimensionality: it is attainable at dimensionality values at most 16 and sustained by a latent asymmetry (i.e., a higher amount) of modalities vs. anti-modalities. In the self-organizing brain, the principle underlies large-scale neural coding for spontaneous behavioral states (movements), on the grounds of deep neural matter-antimatter asymmetry (i.e., of a higher amount of excitatory vs. inhibitory neural subpopulations). See Fig. 1 in [9] for a simple, non-technical explanation of the concept of W-duality allied to *quantum superposition*.

One may associate each vertex $v \in \mathcal{V}$ in $\mathcal{D}_\gamma$ with a *two-level quantum system*, $\mathcal{Q}^v \simeq (\mathfrak{C}^0_{p,q})^2$, or qubit, with computational basis states $|0\rangle^v$ and $|1\rangle^v$ corresponding to the two states which, given the *wave function* $\Box\varphi: \mathcal{V} \to \mathfrak{C}^0_{p,q}$ (as defined in [9]), are implicit in the following condition:

for every $v \in \mathcal{V}$

$$\Box\varphi^*(v) \in \mathcal{O}(\delta) \text{ and, self-dually, } \Box\varphi^*(v) \in \mathcal{O}^0_{p,q}, \quad (1)$$

with respect to the special quadratic algebra $(\mathfrak{C}^0_{p,q}, \delta)$. For $p \neq 0$ (1) defines a *mirror-supersymmetric* flow.

A pure *qubit state* $|\psi\rangle^v$ at $v \in \mathcal{V}$ would then be given by

$$|\psi\rangle^v = \alpha|0\rangle^v + \beta|1\rangle^v, \quad (2)$$

with $\alpha, \beta \in \mathfrak{C}^0_{p,q}$ and $\delta(\alpha) + \delta(\beta) = 1_{\mathfrak{C}^0_{p,q}}$.

We write $\mathcal{Q} \coloneqq \bigotimes_{v \in \mathcal{V}} \mathcal{Q}^v$ for the *quantum vertex module* of $\mathcal{D}$. For an exponentially large network, this treatment would clearly necessitate an exponentially large number of qubits. We therefore opt for a reduced strategy by encoding the vertices of the *geometrical minor* of $\mathcal{D}$ (a population-level structure obtained in analogy to dimensionality reduction [9]) into qubits. The respective crystal architecture may therewith be implemented on a register of at most 16 qubits.

## III. LORENTZ NEURAL CRYSTALS

Fig. 1 illustrates neural crystals classically describable via special Clifford-Lipschitz operations in $\mathfrak{C}^0_{1,3}$ and $\mathfrak{C}^0_{1,4}$, as well as their co-crystals obtained by functional duality. The structures feature precisely two medial layers. As is characteristic of Lorentz neural structures [9], there is a unique lateral layer of exclusively inhibitory interneurons, herein representing anti-modalities. Overall, the graph topology illustrated in Fig. 1(a) (Fig. 1(c)) comprises a Lorentz, (1,3)-partition (resp., a Lorentz, (1,4)-partition): any two distinct deep clusters/subpopulations of neurons of the same layer are necessarily disconnected. The graph topology illustrated in Fig. 1(b) (Fig. 1(d)) comprises a Lorentz, co-(1,3)-partition (resp., a Lorentz, co-(1,4)-partition): any two distinct deep clusters of neurons of the same layer induce a semicomplete subdigraph.

*Remark 1:* While of special interest, Lorentz neural (co-)crystals are not the unique (co-)structures derivable via special Clifford-Lipschitz operations. Depending on values of the pair $(p,q)$, several other (co-)structures are obtainable.

*Remark 2:* Crystals and co-crystals are *coexistable* on the same structure, the composite dynamics being sustained by *hyper-self-duality* or *quantum entanglement* (see Fig. 1 in [9]).

*Remark 3:* For dimensionality values $2 < d < 16$, with $d \neq 4, 8$ a transition from purely algebraic to quasi-spin-geometrical multidimensional patterns of network dynamics is expected.

*Remark 4:* A distinct class of self-organizing networked systems of dimensionality precisely equal to 128 has been alluded to in [9], potentially allied to octonionic dynamics. Rather than subserving behaviors, such dynamics may neurobiologically subserve emotions or internal states and, by projection, be of relevance to far more convoluted analogues in the artificial network sciences.

## IV. A REAL-WORLD SCENARIO

Antagonistic interactions are a common, yet largely ignored hallmark of large-scale, complex networked systems under real-world scenarios: for instance, leading causes of cancer-related deaths are highly heterogeneous, involving almost *non-modifiable* risk factors (socio-demographics, genetic conditions or medical history: modalities), as opposed to almost *modifiable* risk factors (environmental stressors, lifestyle and behavioral risk factors: anti-modalities), the latter being containable by appropriate policy-making and public health interventions [17, 18].

Being considered a marker of socioeconomic development, colorectal cancer (CRC) is the second leading cause of cancer deaths globally, with increasing incidence rates in younger adults mirroring high Socio-demographic Index (SDI) [18]. A neural crystal of dimensionality $d = 5$ for population-level, non-hereditary early-onset CRC (EO-CRC) risk prediction in the European Union (EU) is illustrated in Fig. 2. Given this structure, we propose that faster convergence of network dynamics – assessable by means of the *Cheeger constant* [9] of the underlying digraph – is innately indicative of increased EO-

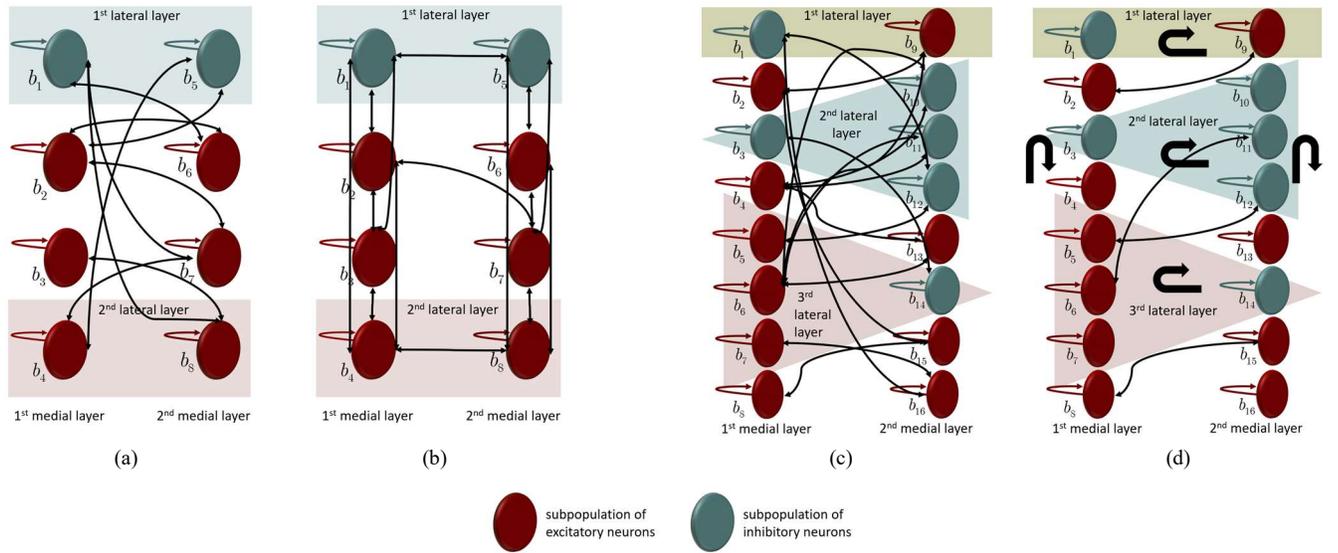

**Fig. 1. Lorentz neural crystals.** (a, c) Spin-geometrical neural structures classically describable via special Clifford-Lipschitz operations in $\mathfrak{C}^{0\times}_{1,3}$ (a) and $\mathfrak{C}^{0\times}_{1,4}$ (c). Either network features precisely one lateral layer of exclusively inhibitory interneurons. Under mirror supersymmetry, the structures accommodate early fusion of 8 (a) and 16 (c) time-invariant (anti-)modalities and the crystallization of latent multidimensional patterns. (b, d) Co-structures with intra-layer connections. Black U-turn arrows in (d) indicate that the respective layers form semicomplete digraphs; $b_i$, $i \in \{1,\ldots,2^{p+q-1}\}$, denotes an element of a basis of $\mathfrak{C}^0_{p,q}$.

CRC incidence. If true, this would further imply that in the EU, sporadic EO-CRC is heterogeneous rather than homogeneous.

Via the crystal the following patterns are recognizable: (i) candidate sporadic EO-CRC risk factors are strongly interrelated (Fig. 2(a)); (ii) the network dynamics are sustained by the self-duality of the socio-demographic modality (Fig. 2(b)), an observation explaining the fact that a high SDI coarsely mirrors increased EO-CRC incidence; (iii) there is a distinctive role for modifiable risk factors (anti-modalities) in multidimensional network dynamics (Fig. 2(c)); (iv) modifiable risk factors are themselves interrelated (Fig. 2(c)). Together, (iii) and (iv) suggest that policy makers may have to prioritize strategies to prevent the adverse interplay of modifiable risk factors in sporadic EO-CRC. Fig. 2(d) presents preliminary outcomes of the neural crystal's implementation on a sample of four EU countries that exhibit significant differences in EO-CRC incidence (45-49y; both sexes), according to the statistics in [18]. The crystal accurately captures the gradient of increasing EO-CRC incidence across the sample, providing early evidence that the disease is at its core heterogeneous.

*Remark:* Explainable MMAI is worth contemplating in relation to other key modelling ingredients, namely *data efficiency* and *human intelligence* (by a human's unique cognitive capacity to leverage causal domain knowledge [19]), the latter being ultimately the essence of human-in-the-loop models.

## V. Outlook

Without blinding to incompleteness [20], we hope to have hereby offered some clues about the feasibility to confront the challenge of interpretability and transparency in MMAI. We attempted to do so by discussing structures standing at the cornerstone of geometric quantum computational properties germane to large-scale neural coding in the self-organizing brain. We alluded to the potential impact of our vision by a specific application to MMAI in healthcare, placing emphasis on an emerging global epidemic: EO-CRC. Future considerations include experimentation with multimodal big data across diverse contexts and model extensions to dynamic analogues – *neural time crystals*. Along these avenues, we aspire to shed light on ongoing debates about deep explainability, chiefly centered around the extent to which human and algorithmic thought are (in)commensurable [21, 22].

### Acknowledgment

We thank Theodore Dalamagas (Athena Research Center, Athens, Greece) for stimulating discussions on the fusion of data-centric with model-centric approaches in explainable AI.

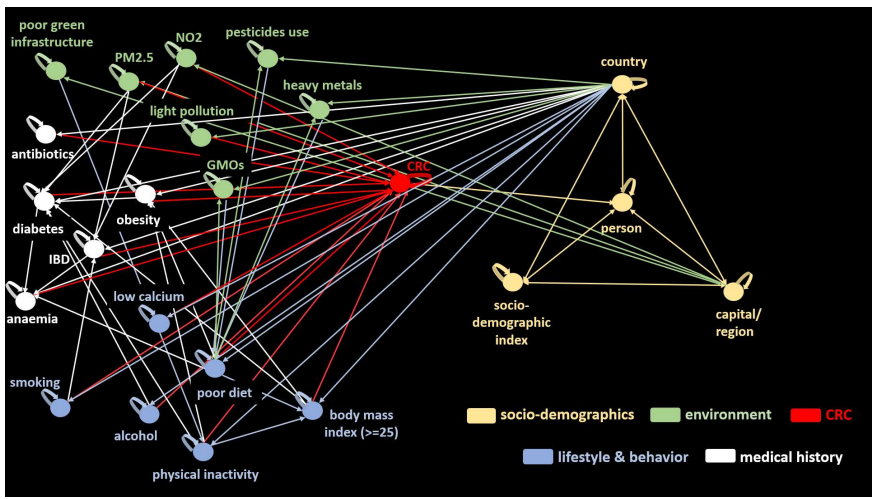 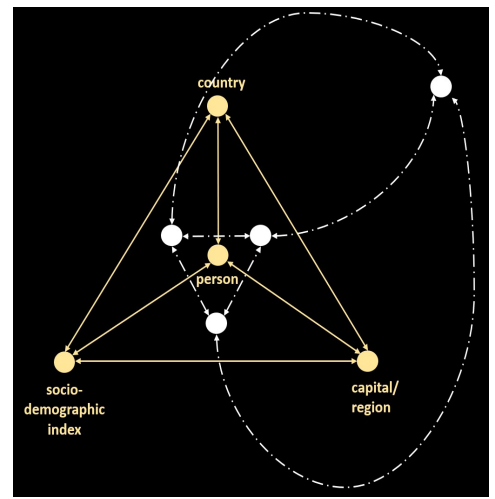

(a) (b)

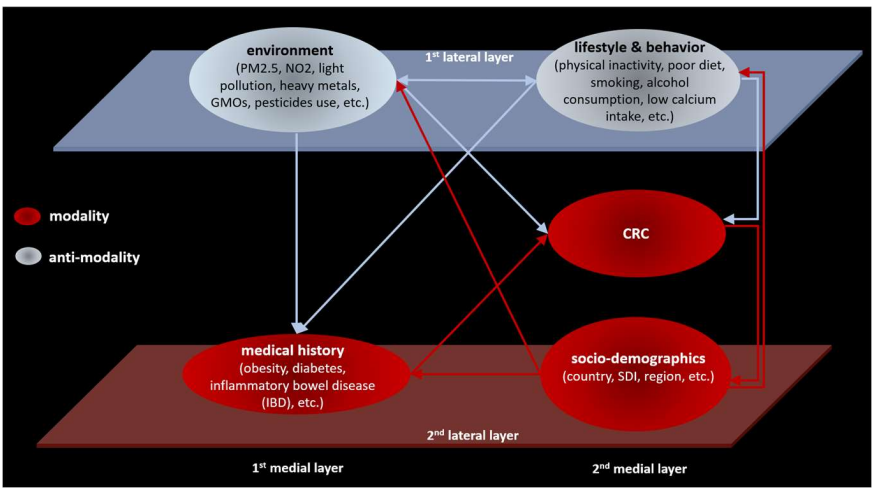 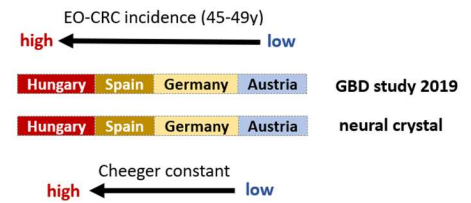

(c) (d)

**Fig. 2. Neural crystal for population-level, sporadic EO-CRC risk prediction in the EU.** (a) Multidimensional (d=5) digraph. Note that the digraph is strongly connected. (b) Contracted (to the modality 'socio-demographics') knowledge digraph (solid lines) and its geometric dual (dashed lines). Observe the self-duality evoking quantum superposition. (c) Geometrical minor of (a) with quasi-spin-geometrical structural characteristics: a quasi-Lorentz, co-(1,3)-partite graph topology, with a unique lateral layer differentiating anti-modalities (modifiable risk factors), which are themselves interrelated. (d) The crystal accurately captures the gradient of increasing EO-CRC incidence (45-49y; both sexes) across a sample of four EU countries, suggesting that the disease is at its core heterogenous. The respective Global Burden of Disease (GBD) Study 2019 [18] and databases therein have been the main source of multimodal data statistics. For clarity of illustration, self-loops in (b) and (c) have been omitted.